\newif\ifpdf					
\ifx\pdfoutput\undefined		
  \pdffalse						
\else							%
  \pdfoutput=1					%
  \pdftrue						%
\fi								%
\documentclass[preprint,showpacs,preprintnumbers,amsmath,amssymb]{revtex4}
\ifpdf
	  \usepackage[pdftex]{graphicx}   
\else
	 \usepackage{graphicx}			  
\fi
	 \usepackage{dcolumn}			  
	 \usepackage{bm}				  
	 
\begin{document}
\title{Quantum optical phenomena in semiconductor quantum dots }
\author{J Thomas Andrews} \email{jtandrews@bitmesra.ac.in}
\affiliation{%
Department of Applied Physics,
Birla Institute of Technology, Mesra, Ranchi 835 215 India}
\begin{abstract}
Quantum optical phenomena are explored 
in artificial atoms well known as semiconductor quantum dots, in the 
presence of excitons and biexcitons.
The analytical results are obtained using the conventional time-dependent 
perturbation technique. Numerical estimations are made for a realistic sample
of CdS quantum dots in a high-Q cavity. Quantum optical phenomena 
such as quantum Rabi oscillations, photon statistics and 
collapse and revival of population inversion in exciton and biexciton states 
are observed. In the presence of biexcitons the collapse and revival 
phenomenon becomes faster due to the strong coupling of biexciton
with cavity field.  
\end{abstract}%
%
\maketitle

\section{Introduction}
Recent progress in crystal growth techniques for the fabrication of 
nanostructure materials has enabled us to realize the
possibility of high-quality epitaxial layers enough to demonstrate
excitonic optical properties of quantum dots (QDs) 
in the ultraviolet \cite{Chen} to infrared \cite{Muka} range   
including room-temperature excitonic lasing, high-temperature
excitonic stimulated emission and high characteristic
temperature for optical threshold power \cite{Bagn,Arak}. Such nonlinear
optical effects would be much more enhanced if
biexcitons were involved in the optical processes because of
the giant oscillator strength effect \cite{Hana,Hohe}. In fact, low-threshold
lasing based on optical processes associated with biexcitons
has been theoretically predicted \cite{Suga,Andr} and observed \cite{Arak}. 

Since its proposal by Purcell \cite{Purc} and early work involving
atoms in cavities \cite{Berm}, cavity quantum electrodynamics
has been actively pursued for its potential
insight into fundamental problems in light-matter interaction.
Here, the energies and modes of photon states, to
which electronic transitions of atoms can couple, are
changed by the cavity confinement. Recently, efforts have
been made to extend this work to the optical regime, which
requires cavities with sizes of the order of few micrometers. This
work has been made possible by the epitaxial fabrication of
high quality layered semiconductors from which optical
cavities have been made that have strong photonic confinement 
in one direction \cite{Weis,Park,Baye}. 
However, the preparation of optical microcavities which confine
the propagation of light in all three dimensions is presently
a challenge for modern microtechnologies. New techniques
are being developed to overcome the difficulties in developing
such microcavities using quantum dot structures.
Quasi atomic light emitters in
these resonators can be mimicked by semiconductor quantum
dots that exhibit discrete density of electronic
states. Arakawa {\it et al},  \cite{Arak} fabricated QD 
lasers with vertically stacked self-assembled
dots with a columnar shape which achieved a room-temperature
threshold current of 5.6~mA, comparable to or even less
than that of quantum-well lasers. The threshold current density
has also been lowered to 90~A/cm$^2$ \cite{Park}. Pulsed and
continuous-wave operation of 1.3~$\mu$m quantum dot lasers
at room temperature has  been realized, which is a milestone towards
the application of QDs to ultrafast fiber optic data transmission and
optical interconnects \cite{Muka}. 
Artemyev {\it et al\/}, \cite{Artm} have developed high-Q microcavities using tiny 
hollow microspheres made of the transparent polymer
polymethylmethacrylate (PMMA) with smooth, highly
reflecting surfaces that exhibit all properties of an optimum optical
cavity. Such improved laser performance has now enabled physicists
to study various unique properties of interaction
between QDs and light.

The study of interaction of quantized radiation with matter is 
important in QDs, because the typical sizes of QDs are 
much smaller than the wavelength of light being used
(for example the typical sizes of QDs are CdS - 1.7nm \cite{Butt}, and
CdSe $\approx$ 4nm \cite{Hess}). The smaller cross sectional areas
of these QDs allow only few photons to
interact with each of them. Hence, the application of quantum optical
tool to QD may yield interesting results.
Accordingly, with the recent developments in the fabrication of high-Q 
microcavities having 3-dimensional photonic confinement on one hand and 
the enhanced excitonic nonlinear optical characteristics of the QDs
in 3-dimensional electronic confinement on the other, 
we have made an attempt to explore the quantum optical phenomena
in these structures in the presence of both excitons and biexcitons. 

\section{Theoretical Calculations}

Time-dependent perturbation technique under the 
interaction picture has been employed to study
the interaction of a single mode of cavity radiation with a
 small QD embedded in a high-Q microcavity.
The atomic-like energy-level structure existing in these QD nanostructures allows
the photo-induced electronic transitions to take place between the ground 
$(|0\rangle)$ and exciton $(|e\rangle)$ states. For large excitation intensities,
this simplistic two-level picture is modified by the creation of biexcitons.
Accordingly, in the present paper, while studying the nonclassical phenomena in QDs,
we have incorporated the exciton and biexciton $(|b\rangle)$ states via a three-level 
system (3-LS) as schematically represented in Fig. 1. The allowed electronic transitions 
(solid arrows) take place between the sates 
$|0\rangle \rightleftharpoons |e\rangle $ and $|e\rangle \rightleftharpoons |b\rangle $.
The parity violating two-photon transitions (dashed arrow) between $|0\rangle \rightleftharpoons |b\rangle$  
are ignored in the present calculations.

For small quantum dots of size ($R$) smaller than the exciton Bohr radius
($a_B$), the energy gap between the vacuum and exciton states is defined
as \cite{And2}  
	\begin{equation}
				\hbar\omega _{oe}=\hbar\omega _g + \frac{\hbar^2}{2 m_r}
				(\frac{\kappa _{ml}}{R})^2, \label{omoe}
	\end{equation} 
where $\hbar \omega _g$ is the band gap energy and $\kappa _{ml}$
is $m$th root of the $l$th order Bessel function with $m$ and $l$
corresponding to the 1$s$, 1$p$, 1$d$, ... 2$s$, 2$p$, 2$d$, ...
levels of the electrons and holes. $m_r$ is the reduced mass of an
electron-hole pair. Due to the finite Coulombic attraction
between two excitons, the energy of the biexciton is reduced by an amount equal 
to the binding energy ($\Delta E$) and defined by \cite{Bany}
   \begin{equation}
				\hbar \omega _{eb} = \hbar \omega _{oe} - \Delta E. \label{omeb}
   \end{equation}

We have assumed that at time $t=0$, all the electrons are in the ground state. 
For time $t > 0$, the interaction of cavity
photons with the QD creates one and two electron-hole pairs (i.e., excitons and
biexcitons). Accordingly, the total wavefunction
of QD-radiation can be defined under dressed state representation as
   \begin{widetext}
   \begin{equation}
				|\Psi(t)\rangle = c_{0, n+1}(t) |0, n+1 \rangle +\sum _{ml} c_{e, n}(t) |e, n \rangle
				+\sum _{ml} c_{b, n-1}(t) |b, n-1 \rangle. \label{psi}
   \end{equation}
   \end{widetext}
$c_{0, n+1}(t)$, $c_{e, n}(t)$ and $c_{b, n-1}(t)$ are the probability amplitudes of the 
vacuum, exciton and biexciton states, respectively. While writing the total
wavefunction, we assume for the field mode  that the initial state 
is coherent, such that 	
	\begin{equation}
				|c_n(0)|^2  = \frac{\bar{n}^n e^{-\bar{n}}}{n!}, 
	\end{equation}
where, $\bar{n}$ is the initial photon number.

The interaction Hamiltonian of the exciton and biexciton states is represented by
   \begin{subequations}
    \begin{equation}
	\label{hami}   					   
				{\cal H}_I = {\cal H}_{e} + {\cal H}_b,
	\end{equation}
with
	 \begin{equation}
				{\cal H}_e = -g_{oe} [a \pi _{eo} e^{-i\Delta _{oe} t} + 
					  	   	 		 	a^\dagger \pi _{oe} e^{i\Delta _{oe} t}]
	\end{equation}
and
	\begin{equation}				
				{\cal H}_b = -g_{eb} [a \pi _{be} e^{-i\Delta _{eb} t} + 
					  	   	 		 	a^\dagger \pi _{eb} e^{i\Delta _{eb} t}],
	\end{equation}
	\end{subequations}
where $a^{\dagger}$ and $a$ are the creation and annihilation
operators of the electromagnetic field, $\pi _{ij}$ $=|i\rangle\langle j|$.
The detuning parameter is defined as $\Delta _{oe(eb)} = 
\omega  - \omega _{oe(eb)}$. $g_{oe} = |\mu _ {e}E/\hbar|\sqrt{n+1}$ 
and  $g_{eb} = |\mu _ {b}E/\hbar|\sqrt{n}$ are the coupling parameters of the 
exciton and biexciton, respectively with the cavity field. Also  
$E$ is the field per photon and
	\begin{subequations}
	\begin{align} 
				 \mu _e = \langle e|\hat{\mu}|0\rangle& = 
				 \mu _{oe}\int\phi(r_e,r)dr   \\
				 \mu _b = \langle b|\hat{\mu}|e\rangle& = 
				 -\sqrt{2}\mu _{oe} \\ 
				 & \times \int\int\int\phi(r_e,r,r_h,r)^* \phi(r_e,r_h)
				 dr\,dr_e\,dr_h.  
	\end{align}
	\end{subequations} 
for the same spin states. \cite{Hu} Here,
$\mu _{oe}=[{e|{p}_{oe}|}/(m_o \omega _{oe})]$ is the
transition dipole moment of the vacuum to exciton
state transition and $m_o$ is  the free electron 
mass. $\phi(r_e,r)$ and $\phi(r_e,r,r_h,r)$ are the exciton and
biexciton wave functions, respectively. They are related to the single
particle wave function $\phi(r)$ as \cite{Bany,Hu} $
\phi(r_e,r)=\phi(r_e)\phi(r)$ and
$\phi(r_e,r,r_h,r)=\phi(r_e)\phi(r)\phi(r_h)\phi(r)$, with \[
\phi(r)=\sqrt{\frac{2a_B^3}{R^3}}
\frac{j_l(\kappa_{nl}r/R)}{j_{l+1}(\kappa_{nl})}.\]
 
Using Schrodinger's equation under the interaction picture, 
the equations of motion of probability amplitudes
are found to be
   \begin{subequations}\label{coeb}
   \begin{align}
				\dot{c}_{0, n+1}(t) = & i\sum _e g_{oe} c_{e, n}(t) \exp(i\Delta _{oe} t), \\
				\dot{c}_{e, n}(t) =& i g_{oe} c_{0, n+1}(t) \exp(-i \Delta _{oe} t)\nonumber \\ 
								  & + i g_{eb} c_{b, n-1}(t) \exp(i\Delta _{eb} t), \\
				\dot{c}_{b, n-1}(t) =& i g_{eb} c_{e, n}(t) \exp(-i\Delta _{eb} t). 
	\end{align}
	\end{subequations}
The above set of coupled equations can be solved by assuming 
   \begin{subequations}\label{cass}
   \begin{align}
				c_{0, n+1}(t) &= A \exp(i\Omega t), \\
				c_{e, n}(t) &= B \exp[i(\Omega - \Delta _{oe}) t], \\
				c_{b, n-1}(t) &= C \exp[i(\Omega - \Delta _{ob}) t], 
	\end{align}
	\end{subequations}
where $A$, $B$ and $C$ are the time-independent constants,
$\Omega$ is an unknown parameter having the dimension of frequency
and $\Delta _{ob} = \Delta _{oe} + \Delta _{eb}$.
Use of (\ref{coeb}) and (\ref{cass}) yields a cubic equation in $\Omega$ as
	\begin{equation}
				\Omega ^3 +a_2 \Omega^2 + a_1^2 \Omega + a_0^3 = 0 \label{Omeg}
	\end{equation}
with $a_0^3 = -|g_{oe}|^2 \Delta _{ob}$; $a_1^2 = \Delta _{oe}\Delta _{ob}
-|g_{oe}|^2 -|g_{eb}|^2$ and $a_2 = -(\Delta _{oe} + \Delta _{ob})$.
The solution of $\Omega$ are found to be
	\begin{subequations} \label{Om123}
    \begin{align}
		\Omega _{_1}  &= -\frac{a_2}{3} + \frac{X^2-12Y^2}{6X},\\
		\Omega _{_3^2}&= -\frac{a_2}{3} - \frac{X^2 e^{\mp i \frac{\pi}{3}}-12Y^2
				e^{\pm i\frac{\pi}{3}}}{6X}.
	\end{align}
	\end{subequations}
with
\begin{widetext}
\[
 X^3=  36a_1^2 a_2-108 a_0^3 -  8 a_2^3+\sqrt{3}\sqrt{
        4 a_1^6 - a_1^4 a_2^2 - 18 a_1^2 a_2 a_0^3+
        27 a_0^6+4 a_0^3 a_2^3 } 
\;\;\; \hbox{and} \;\;\; Y^2=a_1^2-\frac{a_2^2}{3}. \]
\end{widetext}
\subsection{Rabi Oscillations}
For analytical confirmation,
we have reduced the generalized equations obtained for a three-level system (3-LS)
to a conventional two-level atomic system (2-LS). For a 2-LS, in the absence of
biexcitons the coefficients are corrected by substituting $g_{eb} = 0$
 and $\Delta_{ob} =0$ and it is found that $a_0 = 0$, $a_1^2 = -|g_{oe}|^2$ and
$a_2 = -\Delta _{oe}$, which reduces the cubic equation to a quadratic equation 
in $\Omega$. We find that $\Omega _{2}$ vanishes in the absence of biexcitons and the
other two solutions remain finite. 
In a 2-LS, the magnitudes of these eigenvalues decide the frequencies of 
oscillation of the probability of the upper and lower levels
(in the present case between vacuum and exciton states) \cite{Cohe}.
This frequency of oscillation is widely termed as 
Rabi frequency. The three solutions of $\Omega$ obtained in the
present case of a 3-LS also exhibit a similar feature: 
$\Omega _{j}$ obtainable from (\ref{Om123})
may correspond to the Rabi oscillations occurring between the vacuum-exciton 
and exciton-biexciton states. However, in the presence of biexcitons, 
analytical results of $\Omega$ could not be obtained due to the complex 
nature of the three solutions obtained in (\ref{Om123}). Hence, we applied
our analysis to a realistic sample of quantum dot of
CdS embedded in an undamped microcavity. The material parameters of CdS taken from
the experimental paper of Butty {\it et al\/} \cite{Butt} are: $\hbar\omega _g=$2.56eV,
$\Delta E$ = 28meV, $R$ = 1.7nm, $a_B$ = 2.9nm and $m_h/m_e$ = 4.2.
Also, to make the present solutions compatible
with the conventional eigenvalues \cite{Cohe}, we define $\Omega _{j}' = 
\Omega _{j} - a_{2}/3$. The nature of dependence of $\Omega_{j}'$
on exciton detuning parameter $\Delta _{oe}$ are exhibited in Figs. 2 and 3.
In obtaining Fig. 2, we have incorporated the contribution of excitons only, while
in Fig. 3, the contribution of both exciton and biexcitons are
incorporated. The solid lines are obtained in the presence of finite cavity field
and the dashed lines represent the same obtained for $n=0$.
To increase the clarity of the curves, a thicker line has been drawn for 
$\Omega _2'$ while $\Omega _1'$ and $\Omega _3'$ share same line-thickness.
Parts of the dashed lines of $\Omega _{1}'$ (below resonance)
and  $\Omega _{3}'$ (above resonance) are not visible since  
$\Omega _{2}'$ for $n=0$ as well as $n\neq 0$ overlaps with them.
The curves obtained in the absence of biexcitons agree with the conventional
results obtained for a 2-LS system.\cite{Cohe} 
For $n = 0$, all the curves vanish at $\Delta _{oe} = 0$.
However, for $n \neq 0$, the eigenvalues 
show repulsion at $\Delta _{oe} = 0$ by an amount equal to $\pm g_{oe}$. 
The repulsive behavior
of the two eigenvalues is attributed to the finite value of QD-cavity field 
coupling. The phenomenon of repulsion of the eigenvalues are also called as
AC Stark shift or dynamic stark shift since the energy levels show shift
due to the dynamic nature of the amplitude of the radiation. 
As expected from the analytical results for a 2-LS, 
one of the solution $(\Omega _{2} = \Omega _{2}' -a_2/3)$ remains 
unchanged at zero. 

For $n = 0$, all the three curves obtained in the presence of both exciton and
biexciton shown in Fig.~3, display biexcitonic
signatures between $\Delta _{oe} = 0$ and $\Delta _{eb} = 0$.
A close observation of the curves obtained for  $n = 0$ yields the following:  
$\Omega _{1,2,3}$ vanish neither at $\Delta _{oe} = 0$ nor at $\Delta _{eb} = 0$.
However, (i) $\Omega _{1}$ shows its minimum value at 
$\Delta _{oe} +\Delta E/2$ and is repelled from the zero line
by an amount $\Delta E /6$; (ii) for $\Delta _{oe} \geq 0 \leq \Delta _{eb}$, the 
second solution remains constant at $-\Delta E /3$; for $\Delta _{oe} \leq 0$, it steadily
increases to a value of $\Delta E /6$ and again
starts decreasing until $\Delta _{eb}$ becomes equal to zero, 
and (iii) for decreasing detuning parameter, the final solution steadily increases to
$-\Delta E/3$ till $\Delta _{oe} = 0$; further, for $\Delta _{oe} \leq 0 \geq \Delta _{eb}$
 it remains constant at $-\Delta E /3$.
For $\Delta _{eb} \leq 0$, it starts decreasing steadily. 

For $n \neq 0$, dramatic
changes occur. Both $\Omega _{1}'$ as well as $\Omega _{3}'$ get
repelled from $\Omega _{2}'$. In the presence of biexciton, $\Omega _{2}'$ plays
a dominant role by repelling both $\Omega _{1}'$ and $\Omega _{3}'$ for
$n=0$ and $n\neq 0$. This leads to the conclusion that $\Omega _{2}'$ 
couples the biexciton states to the exciton states and hence the ground state. 
In addition to 
the shift along Y-axis, a small shift along X-axis can also be noted 
for all three curves in the presence of biexcitons. For $n \neq 0$ the
number of biexciton becomes finite, such that $a_1$ as defined earlier modifies
the resonance frequency. This leads to small red and blue shifts of the 
peak values of the eigenvalues. These shifts indicate
strong evidence for biexciton signatures in the eigenvalues $\Omega _1$ and $\Omega _2$.
And $\Omega _{3}$ does not show significant change in 
the presence and absence of biexcitons at $\Delta _{oe} =0$ and $\Delta _{eb}=0$.
 
The three solutions of $\Omega$ obtained in (\ref{Om123}) suggest the following
modifications in the assumed solution (\ref{cass})
		\begin{subequations}\label{cfin}
		   \begin{align}
				c_{0, n+1}(t) &= A_1 e^{i\Omega _1 t} + A_2 e^{i\Omega _2 t} + A_3 e^{i\Omega _3 t},\\
				c_{e, n}(t) &= (B_1 e^{i\Omega _1 t} + B_2 e^{i\Omega _2 t} + B_3 e^{i\Omega _3 t}) e^{-i\Delta _{oe}t},\label{coebf}\\
				c_{b, n-1}(t) &= (C_1 e^{i\Omega _1 t} + C_2 e^{i\Omega _2 t} + C_3 e^{i\Omega _3 t}) e^{-i\Delta _{ob}t}.								 					
	\end{align}
	\end{subequations}

In order to calculate the nine time-independent unknown coefficients
$A_i$, $B_i$ and $C_i$ ($i = 1,2,3$), we assume that, at $t = 0$, the QD is 
in the ground state $|0\rangle$ and the cavity field is in a coherent state. 
The corresponding boundary 
conditions at $t=0$, are $c_{0, n+1}(t)=c_n(0)$ 
and $c_{e, n}(t) = c_{b, n-1}(t)=0$. 
Use of these conditions in (\ref{coeb}), (\ref{cass})
and (\ref{cfin}) yields the solution of the unknown coefficients as
	\begin{widetext}
   \begin{equation}\label{ABC}
			   \begin{pmatrix}A_1 & B_1 & C_1 \cr 
			   				  A_2 & B_2 & C_2 \cr 
							  A_3 & B_3 & C_3 \end{pmatrix} = -\frac{c_n(0)}{|M|}
			   \begin{pmatrix}
			   \beta _{23} (\alpha _{23} + |g_{oe}|^2) 		& 
			   		 g_{oe}\beta _{23} (\Delta _{oe}-\gamma _{23}) &
					 	   g_{oe} g_{eb} \beta _{23}	   	\cr
			   \beta _{31} (\alpha _{31} + |g_{oe}|^2) 	   	&
			   		 g_{oe}\beta _{31} (\Delta _{oe}-\gamma _{31}) &	   
					 	   g_{oe} g_{eb} \beta _{31} 		\cr
  		       \beta _{12} (\alpha _{12} + |g_{oe}|^2) 		&
			   		 g_{oe}\beta _{12} (\Delta _{oe}-\gamma _{12}) &
					  	   g_{oe} g_{eb} \beta _{12}\end{pmatrix},
	\end{equation}
	\end{widetext}
with 
$	M = \begin{pmatrix}1 & 1 & 1\cr \Omega_1 & \Omega _2 & \Omega _3 \cr
	 			  \Omega _1^2 & \Omega _2^2 & \Omega _3^2 \end{pmatrix}, $
			   $ \alpha _{ij} = \Omega _{i} \Omega _{j},$ $ \beta _{ij} = \Omega _{i} - \Omega _{j}$
and $ \gamma _{ij} = \Omega _i + \Omega _j.$  
For a 2-LS the above set of equations reduces to 
	\begin{equation}\label{ABC2L}
			   \begin{pmatrix}A_1 & B_1 & C_1 \cr 
			   				  A_2 & B_2 & C_2 \cr 
							  A_3 & B_3 & C_3 \end{pmatrix} = 
					-\frac{c_n(0)}{\Omega _{1} - \Omega _{2}}
					\begin{pmatrix}-\Omega _2 & g_{oe} & 0 \cr 
										   \Omega _1 & -g_{oe} & 0 \cr
										   		  0 & 0 & 0 \end{pmatrix}
		\end{equation}
In obtaining the above equations, we have assumed the biexciton-cavity coupling 
term $(g_{eb})$ to be zero. Comparison of (\ref{ABC2L}) with 
the standard solutions \cite{Cohe} for a 2-LS shows good
agreement and confirms
the validity of the model to a two-level system. 

Using (\ref{cfin})-(\ref{ABC2L}), we have obtained  
the temporal variations of probabilities and exhibited these  
in Fig. 4. The solid, dashed and dotted lines represent
$|c_{o,n-1}|^2$, $|c_{e,n}|^2$ and $|c_{b,n+1}|^2$,
respectively.  The inset shown in the figure demonstrates the same parameters in 
the absence of biexciton contribution. In the absence of biexciton, the
ground and exciton state populations execute oscillation at a frequency
equal to $2g_e$. Hence, the ground state is bleached for $g_e t = (2N+1)\pi/2$ with
$N = 0, 1, 2, ...$ . As expected from earlier analytical simplifications, $|c_{b,n+1}|^2$
remains constant at zero.
In the presence of biexciton contribution,
the boundary conditions are exactly followed. For $t>0$,
the ground state population starts decreasing until $g_e t = \pi/2$.
Simultaneously, the exciton population reaches its maximum when $g_e t = \pi/4$,
and later decreases to a minimum value
at $g_e t = \pi/2$. However, the biexciton population increases
very slowly but attains its maximum value when $g_e t = \pi/2$.
This region is marked as \textbf{1} in the figure. This is an  
interesting region, since the population between $|b\rangle$
and $|0\rangle$ as well as between $|b\rangle$ and $|e\rangle$ 
is inverted. This population
inverted region occurs again when $g_e t = 5\pi/2$. 
The significance of the region \textbf{1} is
that for $g_e t = (5N+1)\pi/2$ and $(N+1)5\pi/2$, low
threshold for exciton and biexciton lasing can be achieved.
Another region of similar interest is 
marked as \textbf{2}, where population inversion
occurs between $|b\rangle$ and $|e\rangle$. All
the three curves rephase for $g_b t = 4N \pi$. 

\subsection{Photon Statistics}
The expressions (\ref{cfin}) represent the probability amplitudes
of the electron to
occupy the states $|0\rangle$, $|e\rangle$ and $|b\rangle$ and the probability of $n$ photons to occupy the states
$|n\rangle$ at any arbitrary time $t$. The trace over the electronic states yield 
the probability $p(n,t)$ of the cavity field as
	\begin{equation}
				p(n,t) =\sum_{ml}\left[{|c_{0, n}(t)|^2 + |c_{e, n}(t)|^2 + |c_{b, n}(t)|^2}\right]. \label{pn}
\end{equation} 
Comparison of the above results with the semiclassical \cite{Brewer}
results show disagreement, since the semiclassical techniques assume 
the photon number distribution 
at $t=0$ as $c_{0,n+1} = 1$. However, for
an initially coherent state, the Poisson distribution allows the field to 
propagate in a coherent manner.
In Fig. 5, variation of $p(n,t)$ with time as well 
as number of photons is demonstrated for $\bar{n} = 10$. 
Fig. 5a and 5b display contour plots in presence
of excitons only and in presence of exciton as well as biexciton, respectively. 
The peak value is found to be at $\bar{n}$. 
Both the curves strictly follow Poisson statistics with respect to $n$,
but the envelope oscillates with time.  
A few peaks are obtained for smaller values of $n$ while the number of peaks starts increasing with
$n$. The reason for these oscillations are attributed to the definition of $g_{0e}$
which is proportional to $\sqrt{n+1}$. 
In the presence of excitonic contribution only, the photon statistics oscillates
at a frequency $g_{0e}$. In the presence of biexcitons a clear case of chaos is observed as evident 
from Figs. 5c and 5d, which are obtained for $n=10$. 
\subsection{Collapse and revival phenomena}
The phenomenon of collapse and revival of Rabi oscillations is studied in
atomic systems using atomic inversion operator. The reason for the
collapse and revival phenomenon is well understood now. The contributions 
corresponding to different $n$'s interfere in such a manner that they
initially go out of phase. After that they acquire a common phase, and 
this process is continuously repeated to obtain a series of collapses and revivals.
In the present generalized model, we have assumed a three-level structure of excitons and
biexcitons. The corresponding population inversion in these states 
can be found from
	\begin{subequations}\label{W}
			\begin{equation}
				W(t)_{0e} = \sum _{e,n}(|c_{0, n+1}(t)|^2 - |c_{e, n}(t)|^2)
			\end{equation}			
and
			\begin{equation}
				W(t)_{eb} = \sum _{e,n}(|c_{e, n}(t)|^2 - |c_{b, n-1}(t)|^2). 
			\end{equation}							
	\end{subequations}
In Fig. 6, the temporal variation of $W_{oe}$ is demonstrated in the presence (curve $a$)
and absence (curve $b$) of biexciton contribution. 
Both the curves exhibit collapse and revival of 
population inversion between the ground and exciton states. A very slow revival is 
noted in the absence of biexcitons. In the presence of biexcitons, the strong
coupling of the biexciton with cavity field reduces the revival time; hence
a very small collapse period is observed. However, in the absence of biexcitons
the QD behaves like a 2-LS; hence  an atomic like collapse and revival
phenomenon is noted. In the absence of the biexciton contribution, 
$W_{0e}$ oscillates around zero. With the biexcitonic contribution, it shows
a positive DC shift. The transient behavior of excitonic and biexcitonic 
inversions, as exhibited in Fig. 7, does not show any remarkable change.
In the presence of the biexciton $W_{0e}$, as well as $W_{eb}$ display the same
collapse and revival time. However, $W_{0e}$ oscillates around 0.3,
while $W_{eb}$ oscillates around -0.2.  
  
\section{Conclusions}
A simple quantum optical model has been developed to 
study nonclassical phenomena in semiconductor
quantum dots in an undamped cavity. On reduction of the present calculations to a 
two-level system, these agree well with the conventional results. 
The results are applied to a realistic 
semiconductor QD of CdS of size 1.7nm. The transient nature of photon statistics 
exhibits oscillations. The collapse and revival phenomenon in the presence of 
biexcitons becomes faster due to the strong biexciton-photon coupling.   

\acknowledgements{The author acknowledges the constant encouragement and 
support received from Dr. Pratima Sen, Prof. P. K. Sen and Prof. P. K. Barhai. 
The author is also thankful to Prof. R. K. Popli for reading the manuscript.}


\newpage

\begin{center}
\bf FIGURE CAPTIONS\/
\end{center}

\begin{itemize}
\item[FIG. 1]  Schematic diagram of three-level structure of ground, exciton and 
					 biexciton states in a quantum dot. Solid lines represent allowed electronic 
					 transitions, while the two-photon transition shown by the dashed line is not allowed.
\item[FIG. 2]  The nature of dependence of eigenvalues $\Omega _{j}'$
				 	  on exciton detuning parameter $\Delta _{0e}$ in CdS quantum dot in the absence
					  of biexciton contribution. 
\item[FIG. 3] Variation of $\Omega _{j}'$ with detuning parameter
		  			 in the presence of biexciton contribution in a small quantum dot of CdS.
\item[FIG. 4] Temporal nature of probabilities of ground (dotted),
		  			 exciton (dashed) and biexciton (solid) states in the presence of exciton and
		  			 biexciton. The inset shows the same in the presence of exciton only. In the regions marked
		  			 as \textbf{1}, population inversion occurs between $|b\rangle$ and $|e\rangle$ as well
		  			 as between $|b\rangle$ and $|e\rangle$. In the regions \textbf{2}, inversion
		  			 occurs between $|b\rangle$ and $|e\rangle$ states. 
\item[FIG. 5] Photon statistics $p(n,t)$ in a small quantum dot of CdS
					 embedded in a microcavity for $\bar{n}=10$. 
		  			 The contour plot $a$ is obtained in the absence of biexcitons while $b$
		  			 is obtained in the presence of biexcitons. Plots (c) and (d) are obtained
		  			 for $n=10$. 
\item[FIG. 6] The temporal variation of excitonic inversion 
	      			 ($W_{oe}$) in a QD of CdS. Curve $a$ is obtained in the presence 
		  			 of biexcitons while curve $b$ is obtained in the absence of biexcitons.
\item[FIG. 7] The nature of dependence of exciton and biexciton
		  			 population inversion ($W_{0e}$ and $W_{eb}$) on time in a small quantum
		  			 dot of CdS.
\end{itemize}

\newpage
\begin{figure}
\vspace{9cm}
\ifpdf
	  	  \includegraphics[width=60mm,angle=0]{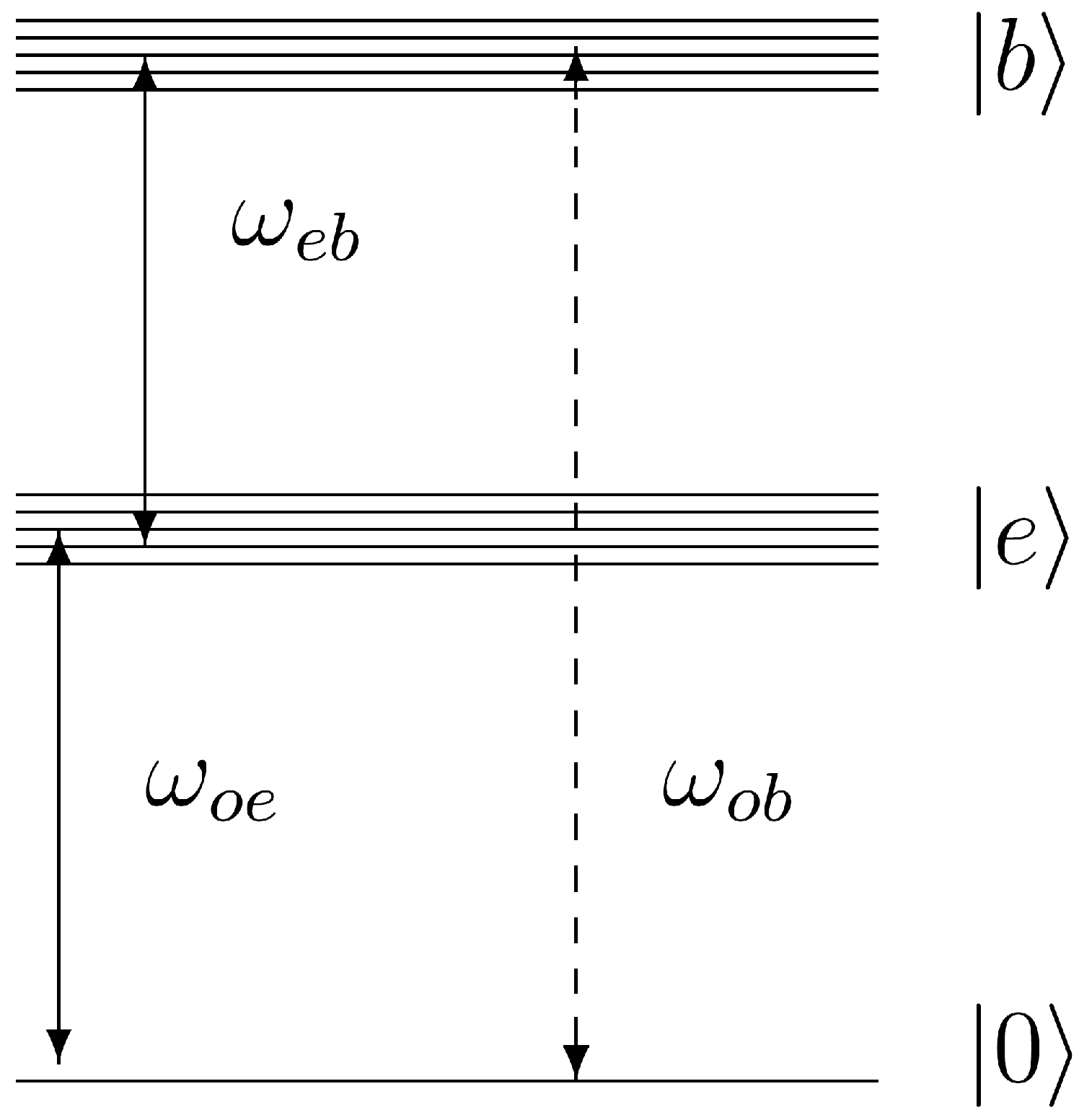}
	  \else 
	  	  \includegraphics[width=60mm,height=60mm]{fig1.bmp}
\fi
\caption{Schematic diagram of three-level structure of ground, exciton and 
					 biexciton states in a quantum dot. Solid lines represent allowed electronic 
					 transitions, while the two-photon transition shown by the dashed line is not allowed.}
\end{figure}  
\hfill
\vfill

\newpage
   		  \begin{figure}
		  \vspace{9cm}
   			  \ifpdf
	  	  \includegraphics[width=80mm,angle=0]{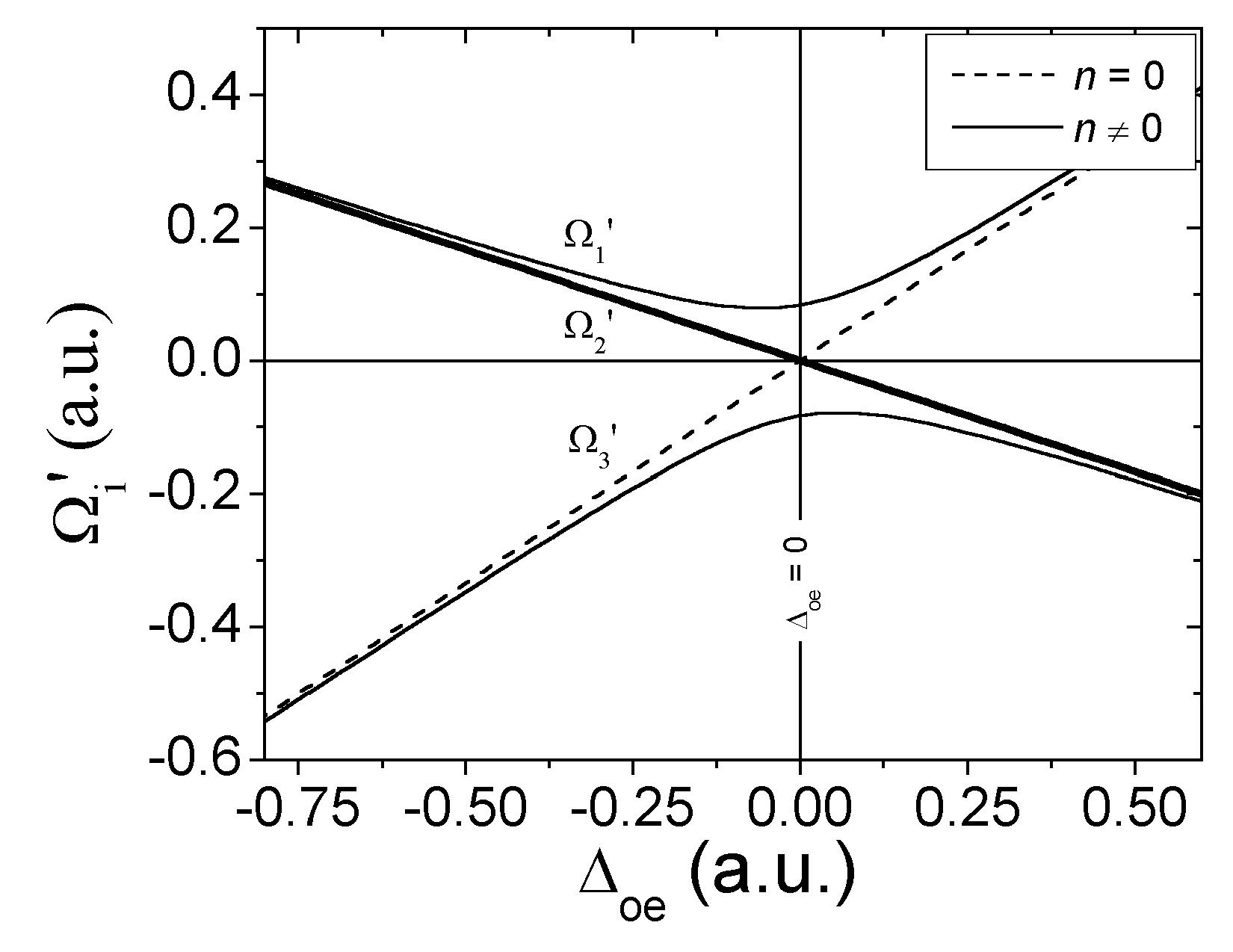}
		  	  \else 
	  	  \includegraphics[width=80mm,height=60mm]{fig2.bmp}
		  		\fi
				\caption{The nature of dependence of eigenvalues $\Omega _{j}'$
				 	  on exciton detuning parameter $\Delta _{0e}$ in CdS quantum dot in the absence
					  of biexciton contribution.}
		  \end{figure}
\hfill

\newpage

		    \begin{figure}
			\vspace{9cm}
   			  \ifpdf
	  	  \includegraphics[width=80mm,angle=0]{fig3.png}
		  	  \else 
	  	  \includegraphics[width=80mm,height=60mm]{fig3.bmp}
		  		\fi
				\caption{Variation of $\Omega _{j}'$ with detuning parameter
		  			 in the presence of biexciton contribution in a small quantum dot of CdS.}
		  \end{figure}
\hfill

\newpage

		  
		    \begin{figure}
   			\vspace{9cm}
			  \ifpdf
	  	  \includegraphics[width=80mm,angle=0]{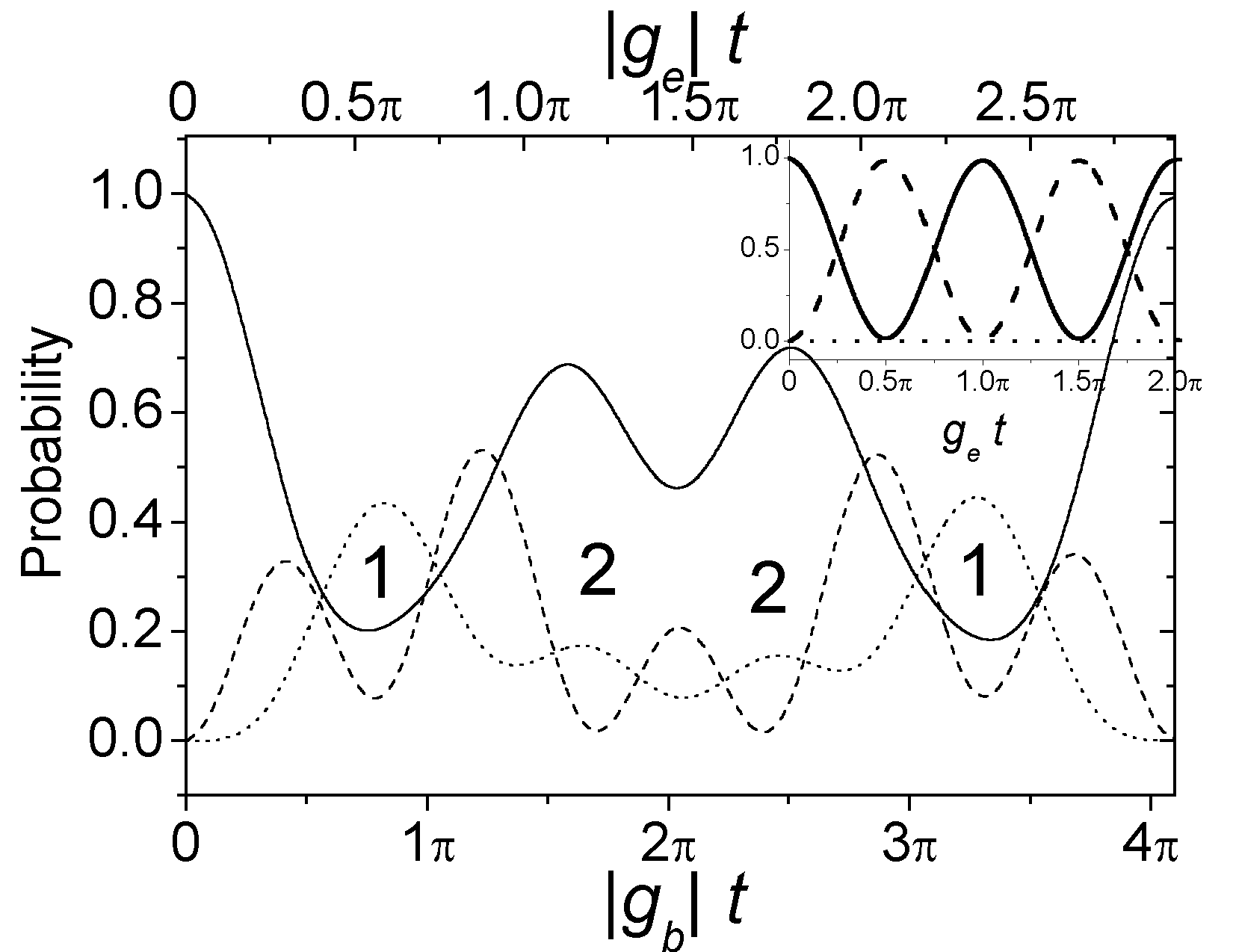}
		  	  \else 
	  	  \includegraphics[width=80mm,height=60mm]{fig4.bmp}
		  		\fi
				\caption{Temporal nature of probabilities of ground (dotted),
		  			 exciton (dashed) and biexciton (solid) states in the presence of exciton and
		  			 biexciton. The inset shows the same in the presence of exciton only. In the regions marked
		  			 as \textbf{1}, population inversion occurs between $|b\rangle$ and $|e\rangle$ as well
		  			 as between $|b\rangle$ and $|e\rangle$. In the regions \textbf{2}, inversion
		  			 occurs between $|b\rangle$ and $|e\rangle$ states.}
		  \end{figure}

\hfill

\newpage

		  	    \begin{figure}
				\vspace{9cm}
   			  \ifpdf
	  	  \includegraphics[width=80mm,angle=0]{fig5.png}
		  	  \else 
	  	  \includegraphics[width=80mm,height=60mm]{fig5.bmp}
		  		\fi
				\caption{Photon statistics $p(n,t)$ in a small quantum dot of CdS
					 embedded in a microcavity for $\bar{n}=10$. 
		  			 The contour plot $a$ is obtained in the absence of biexcitons while $b$
		  			 is obtained in the presence of biexcitons. Plots (c) and (d) are obtained
		  			 for $n=10$.}
		  \end{figure}
\hfill

\newpage
	    \begin{figure}
		\vspace{9cm}
   			  \ifpdf
	  	  \includegraphics[width=80mm,angle=0]{fig6.png}
		  	  \else 
	  	  \includegraphics[width=80mm,height=60mm]{fig6.bmp}
		  		\fi
				\caption{The temporal variation of excitonic inversion 
	      			 ($W_{oe}$) in a QD of CdS. Curve $a$ is obtained in the presence 
		  			 of biexcitons while curve $b$ is obtained in the absence of biexcitons.}
		  \end{figure}

\hfill

\newpage
	  	    \begin{figure}
			\vspace{9cm}
   			  \ifpdf
	  	  \includegraphics[width=80mm,angle=0]{fig7.png}
		  	  \else 
	  	  \includegraphics[width=80mm,height=60mm]{fig7.bmp}
		  		\fi
				\caption{The nature of dependence of exciton and biexciton
		  			 population inversion ($W_{0e}$ and $W_{eb}$) on time in a small quantum
		  			 dot of CdS.}
		  \end{figure}

\end{document}